\documentclass[twocolumn,preprintnumbers,amsmath,amssymb,aps,nobibnotes,nofootinbib]{revtex4} %superscriptaddress
\pdfoutput=1

\usepackage{graphicx}
\graphicspath{{Figs/}}
\usepackage{dcolumn}% Align table columns on decimal point
\usepackage{bm}% bold math
\usepackage[a4paper, top=1.0in, bottom=1.2in, left=1.0in, right=1.0in]{geometry}
\usepackage{multirow}

\renewcommand{\>}{\rangle}

\newcommand{\beqn}{\begin{eqnarray}}
\newcommand{\eeqn}{\end{eqnarray}}
\def\sla#1{\setbox0=\hbox{$#1$}\dimen0=\wd0
      \setbox1=\hbox{/} \dimen1=\wd1 \ifdim\dimen0>\dimen1
      \rlap{\hbox to \dimen0{\hfil/\hfil}} #1                        \else
      \rlap{\hbox to \dimen1{\hfil$#1$\hfil}}
      /   \fi}

\newcommand{\nn}{\nonumber}
\newcommand{\ov}{\overline}
\newcommand{\al}{\alpha}

\newcommand{\ga}{\gamma}

\newcommand{\D}{\Delta}
\newcommand{\eps}{\epsilon}

\newcommand{\la}{\lambda}

\newcommand{\bk}{\hat B_K}
\newcommand{\keps}{\kappa_{\epsilon}}
\newcommand{\ovkeps}{{\overline\kappa}_{\epsilon}}

\newcommand{\mc}{\mathcal}

\long\def\symbolfootnote[#1]#2{\begingroup%
 \def\thefootnote{\fnsymbol{footnote}}\footnote[#1]{#2}\endgroup}
\newcommand{\mysection}[1]{\section{\boldmath #1}}
\newcommand{\mysectionA}[1]{\section*{\boldmath #1}}

\newcommand{\lc}{\lowercase}
\newcommand{\noi}{\noindent}

\begin{document}

\preprint{TUM-HEP-688/08}

\title{\boldmath Correlations among new CP violating effects in $\D F = 2$ observables}

\author{Andrzej~J.~Buras}
\affiliation{Physik-Department, Technische Universit\"at M\"unchen, D-85748 Garching, Germany}

\author{Diego~Guadagnoli}
\affiliation{Physik-Department, Technische Universit\"at M\"unchen, D-85748 Garching, Germany}

\date{\today}

\begin{abstract}

\noi We point out that the observed CP violation in $B_d - \ov B_d$ mixing, taking into account the measured ratio 
$\D M_d / \D M_s$, the recently decreased lattice value of the non-perturbative parameter $\bk$ and an additional 
effective suppression factor $\keps \simeq 0.92$ in $\eps_K$ neglected sofar in most analyses, may be insufficient 
to describe the measured value of $\eps_K$ within the Standard Model (SM), thus hinting at new CP violating contributions
to the $K - \ov K$ and/or $B_d - \ov B_d$ systems. Furthermore, assuming that $\D M_d / \D M_s$ is SM-like,
the signs and the magnitudes of new physics effects in $\eps_K$ and in the CP asymmetries $S_{\psi K_s}$ and $S_{\psi \phi}$
may turn out to be correlated. For example, in a scenario with new CP-phases in $B_d$ and $B_s$ mixings being approximately 
equal and negative, a common new phase $\approx - 5^\circ$ could remove the tension between $\eps_K$ and $S_{\psi K_s}$
present in the SM and simultaneously accommodate, at least partly, the recent claim of $S_{\psi \phi}$ being much larger 
than the SM expectation.
We emphasize the importance of precise determinations of $V_{cb}$, $\bk$, $F_K$ and $\xi_s$,~to which the parameter 
$\eps_K$ and its correlation with the CP violation in the $B_d - \ov B_d$ system are very sensitive.

\end{abstract}

%\pacs{}

\maketitle

\mysection{I\lc{ntroduction}}\label{sec:intro}

\noi The major task achieved in quark flavour physics up to the present is a sound test of the Standard 
Model (SM) mechanism of flavour and CP violation. This mechanism has proven to be able to accommodate dozens of 
measured processes, to a degree of accuracy sometimes unexpected. These processes have consequently allowed a 
redundant determination of the CKM matrix parameters, in particular $\ov \rho, \ov \eta$. Indeed, the $(\ov \rho, 
\ov \eta)$-plots by the UTfit and CKMfitter collaborations have become somewhat an icon of the SM performance in 
flavour physics. To the present level of accuracy, the `big picture' in flavour and CP violation looks therefore 
quite solid.

Nonetheless, hints of discrepancies with respect to the SM expectations do exist in some flavour observables. The 
most recent is the claim of a $B_s$ mixing phase much larger than the SM prediction. This conclusion --~first 
signalled in 2006 by Lenz and Nierste \cite{LenzNierste}~-- has been recently reported as an evidence
by the UTfit collaboration \cite{UTfit-Bs} on the basis of a combined fit to the time-dependent tagged angular analyses of 
$B_s \to \psi \phi$ decays by the CDF \cite{CDF-tagged} and D{\O} \cite{D0-tagged} collaborations. 
The result of \cite{UTfit-Bs} urges higher-statistics data from Tevatron, but, if confirmed, would be the first 
evidence of physics beyond the SM from collider data.

Another emblematic example, also emphasized in \cite{UTfit-Bs,LunghiSoni-Bd}, is that of the penguin-dominated 
non-leptonic $b \to s$ decays. The mixing-induced CP asymmetries measured in these decays allow to access 
$\sin 2 \beta$, where $\beta$ is one of the angles of the Unitarity Triangle (UT), defined below in 
eq. (\ref{betadefs}). The $\sin 2 \beta$ determinations obtained from these decay modes are 
systematically lower than the value measured in the tree-level decay $B_d \to \psi K_s$. The latter 
direct determination has in turn been found to be lower than the one extracted indirectly 
from tree-level measurements, in particular $|V_{ub}/V_{cb}|$ \cite{UTfit-phid,BBGT,BallFleischer}. Conclusions in 
this respect depend mostly on the $|V_{ub}|$ estimate, which is a not yet settled issue.
Independently of this, the problem has been recently revived in \cite{LunghiSoni-Bd} as a consequence of a new lattice 
estimate of the $\hat B_K$ parameter \cite{DJAntonio}, which reads: $\bk = 0.720(13)(37).$\footnote{Similar results have 
been obtained in \cite{JLQCD-BK}, while $\bk = 0.83(18)$ has been reported in \cite{HPQCD-BK}. It may also be interesting to 
note that some non-lattice estimates of $\bk$, e.g. those in the large $N_c$ approach, feature $\bk \lesssim 0.70$. See in 
particular refs. \cite{BardeenBurasGerard-BK,PichDeRafael-BK,BijnensPrades-BK}.} 
The parameter $\bk$ enters the CP-violating observable $\eps_K$ and, in the context of the SM, the decrease of $\bk$ found 
in \cite{DJAntonio,JLQCD-BK} with respect to previous determinations favors $\sin 2 \beta$ again substantially higher than 
the one extracted from $B_d \to \psi K_s$.

Here we would like to gather these pieces of information and try to address the question whether existing 
data on the $B_d$ and $K$ systems do already signal the presence of inconsistencies in the SM picture
of CP violation from a somewhat different point of view than the analysis in \cite{LunghiSoni-Bd}. 
More concretely, the most updated theoretical input in $K$ physics --~in particular the quite 
low central value from the aforementioned new lattice determination of $\hat B_K$ and an additional effective 
suppression factor $\keps \simeq 0.92$ in the SM $\eps_K$ formula neglected in most analyses to date~-- tend both to lower 
the SM prediction for $|\eps_K|$ beneath its measured value if the amount of CP violation in the $B_d$ system, 
quantified by $\sin 2\beta$ from $B_d \to \psi K_s$, is used as input.
 
In order to cure this potential inconsistency, one should then introduce either a new CP phase in the $B_d$ or respectively 
in the $K$ system, or alternatively two smaller phases in both systems. The case of a single additional $B_d$ mixing phase 
is especially interesting. In this instance, the SM formula for the mixing-induced CP asymmetry $S_{\psi K_s}$ generalizes to
\beqn
S_{\psi K_s} = \sin(2 \beta + 2 \phi_d) = 0.681 \pm 0.025~,
\label{SpsiKs}
\eeqn
where $\phi_d$ is the new phase. The information mentioned above points toward a small {\em negative} value of $\phi_d$. 
On the other hand, the mixing-induced CP asymmetry $S_{\psi \phi}$ is given by \cite{BBGT}
\beqn
S_{\psi \phi} = \sin(2 |\beta_s| - 2 \phi_s)~,
\label{Spsiphi}
\eeqn
where the SM phases $\beta, \beta_s$ are defined from the CKM matrix entries $V_{td}, V_{ts}$ through
\beqn
V_{td} = |V_{td}| e^{- i \beta}~, ~~~ V_{ts} = - |V_{ts}| e^{- i \beta_s}~,
\label{betadefs}
\eeqn
with $\beta_s \approx - 1^\circ$. From eq. (\ref{Spsiphi}) one finds that a negative $\phi_s$ is also required 
to explain the claim of \cite{UTfit-Bs}. It is then tempting to investigate whether, at least to first approximation, the same 
new phase $\phi_d \approx \phi_s \approx \phi_B$ could fit in both $B_d$ and $B_s$ systems, being a small correction 
in the former case --~where the SM phase is large~-- and the bulk of the effect in the latter.\footnote{This simple correlation 
is unrelated to more involved correlations that invoked $\D F = 1$ transitions, as in \cite{BFRS} and references therein.}

The rest of this paper is an attempt to explore the above possibilities in more detail. For the sake of clarity, we introduce 
here some notation details. The amplitudes for $B_q$ ($q = d, s$) meson mixings are parameterized as follows
\beqn
\<B_q | \mc{H}^{\rm full}_{\D F = 2} | \ov{B}_q \> &\equiv& A_q^{\rm full} e^{2 i \beta_q^{\rm full}}~,
\label{parameterization}
\eeqn
where, to make contact with the conventions on the SM phases $\beta, \beta_s$, one has
\beqn
\beta_d^{\rm full} &=& \beta + \phi_d~, \nn \\
\beta_s^{\rm full} &=& \beta_s + \phi_s~.
\label{betadsdefs}
\eeqn
The magnitudes $A_q^{\rm full}$ can be written as
\beqn
&&\hspace{-0.2cm}A_q^{\rm full} = A_q^{\rm SM} C_q~, \nn \\
&&\hspace{-0.2cm}\mbox{with  }A_q^{\rm SM} \equiv |\<B_q | \mc{H}^{\rm SM}_{\D F = 2} | \ov{B}_q \>| = \D M_q^{\rm SM}/2~.~~~~~
\label{SMdefs}
\eeqn
Concerning $C_q$, with present theoretical errors on the $B_q$ system mass differences $\D M_q$, it is impossible to draw 
conclusions on the presence of NP. Therefore, one typically considers the ratio $\D M_d / \D M_s$, where the theoretical error is 
smaller, and is dominated by the uncertainty in the lattice parameter $\xi_s$, defined as
\beqn
\xi_s \equiv \frac{F_{B_s} \sqrt{\hat B_s}}{F_{B_d} \sqrt{\hat B_d}}~.
\label{xis}
\eeqn
The resulting SM prediction for $\D M_d / \D M_s$ is in good agreement with the experimentally 
measured ratio\footnote{Variations of the SM formula due to different CKM input are much smaller than the relative theoretical 
error, which is roughly 2$\times \sigma_{\xi_s}$.}. Hence it is plausible, at least to first approximation, to assume 
$\D M_d / \D M_s$ as unaffected by NP, i.e., recalling eq. (\ref{SMdefs}), that
\beqn
C_d = C_s = C_B~.
\label{Cd=Cs}
\eeqn
We will comment on this assumption later on in the analysis.

\mysection{$\eps_K$ \lc{and $\sin 2 \beta$}}\label{sec:K}

\noi We start our discussion by looking more closely at the $\eps_K$ parameter. For the latter, we use the following 
theoretical formula \cite{Uli-private}
\beqn
&&\eps_K = e^{i \phi_\eps} \sin \phi_\eps \left( \frac{{\rm Im}(M^K_{12})}{\D M_K} + \xi \right)~,\nn \\
&&\xi = \frac{{\rm Im} A_0}{{\rm Re} A_0}~,
\label{epsexact}
\eeqn
with $A_0$ the 0-isospin amplitude in $K \to \pi \pi$ decays, $M^K_{12} = \<K | \mc{H}^{\rm full}_{\D F = 2} | \ov K \>$ and 
$\D M_K$ the $K - \ov K$ system mass difference. The phase $\phi_\eps$ is measured to be \cite{PDG}
\beqn
\phi_\eps = (43.51 \pm 0.05)^\circ~.
\eeqn
Formula (\ref{epsexact}) can for instance be derived from any general discussion of the $K - \ov K$ system 
formalism, like \cite{Chau-BSS,BurasLesHouches}, and can be shown to be equivalent to eq. (1.171) of \cite{FNAL-report-2002}, 
where all the residual uncertainties are explicitly indicated and found to be well below 1\%. In contrast with the $\eps_K$ formula 
used in basically all phenomenological applications, eq. (\ref{epsexact}) takes into account $\phi_\eps \neq \pi/4$ and $\xi \neq 0$. 
Specifically, the second term in the parenthesis of eq. (\ref{epsexact}) constitutes an O(5\%) correction 
to $\eps_K$ and in view of other uncertainties was neglected until now in the standard analyses of the UT, with the notable exception 
of \cite{AOV-epsK,AOV-BK}. Most interestingly for the discussion to follow, both $\xi \neq 0$ and $\phi_\eps < \pi/4$ imply 
suppression effects in $\eps_K$ relative to the approximate formula. In order to make the impact of these two corrections transparent, 
we will parameterize them through an overall factor $\kappa_\eps$ in $\eps_K$:
\beqn
\keps = \sqrt 2 \sin\phi_\eps \ovkeps~,
\eeqn
with $\ovkeps$ parameterizing the effect of $\xi \neq 0$. The calculation by Nierste in \cite{FNAL-report-2002} (page 58), the analyses 
in \cite{AOV-epsK,AOV-BK} and our very rough estimate at the end of the paper show that $\ovkeps \lesssim 0.96$, with $0.94\pm 0.02$ 
being a plausible figure. Consequently we find
\beqn
\keps = 0.92 \pm 0.02~.
\label{keps}
\eeqn
In view of the improvements in the input parameters entering $\eps_K$, the correction (\ref{keps}) may start having a non-negligible 
impact in UT analyses. Therefore, a better evaluation of this factor would certainly be welcome.

One can now identify the main parametric dependencies of $\eps_K$ within the SM through the formula
\beqn
&&\hspace{-0.4cm} |\eps_K^{\rm SM}| = \keps C_\eps \bk |V_{cb}|^2 \la^2 \ov \eta \times \nn \\
&& \left( |V_{cb}|^2 (1- \ov \rho) \eta_{tt} S_0(x_t) + \eta_{ct} S_0(x_c,x_t) - \eta_{cc} x_c \right)~, \nn \\
[0.2cm]
&&\hspace{-0.4cm} \mbox{with  } C_\eps = \frac{G_F^2 F_K^2 m_{K^0} M_W^2}{6 \sqrt 2 \pi^2 \D M_K}~,
\label{epsapprox}
\eeqn
and where notation largely follows ref. \cite{BurasLesHouches}, in particular $x_i = m_i^2(m_i)/M_W^2$, $i = c,t$. 
As far as CKM parameters are concerned, eq. (\ref{epsapprox}) reproduces the `exact' SM result, where 
no expansion in $\la$ is performed, to 0.5\% accuracy. Now, $1- \ov \rho = R_t \cos \beta$ and $\ov \eta = R_t \sin \beta$, 
where the UT side $R_t$ is given by
\beqn
R_t &\approx& \frac{1}{\la} \frac{|V_{td}|}{|V_{ts}|}
\nn \\
&=& \frac{\xi_s}{\la} \sqrt{\frac{M_{B_s}}{M_{B_d}}} \sqrt{\frac{\D M_d}{\D M_s}} \sqrt{\frac{C_s}{C_d}}~.
\label{Rt}
\eeqn
with $C_d = C_s$ assumed here (see eq. (\ref{Cd=Cs})) and $\xi_s$ introduced in eq. (\ref{xis}).
Therefore, for the leading contribution to $\eps_K$, due to top exchange, one can write
\beqn
|\eps_K| \propto \keps F_K^2 \bk |V_{cb}|^4 \xi_s^2 \frac{C_s}{C_d} \sin 2 \beta~,
\label{epsprop}
\eeqn
showing that the prediction for $\eps_K$ is very sensitive to the value of $|V_{cb}|$ but also 
to $\xi_s$ and $F_K$. All the input needed in eqs. (\ref{epsapprox})-(\ref{epsprop}) and in the rest of our paper is reported 
in table \ref{tab:input}.
\begin{table}[th]
\footnotesize
\center{
\begin{tabular}{|l|l|}
\hline
& \\
[-0.25cm]
$G_F = 1.16637 \cdot 10^{-5}$ GeV$^{-2}$ & $\lambda = 0.2255(7)$ \hfill \cite{Flavianet} \\
$M_W = 80.403(29)$ GeV & $|V_{cb}| = 41.2(1.1) \cdot 10^{-3}$ \hfill \cite{PDG08}\\
\cline{2-2}
& \\
[-0.25cm]
$M_Z = 91.1876(21)$ GeV & $\eta_{cc} = 1.43(23)$ \hfill \cite{HerrlichNierste} \\
$\al_s(M_Z) = 0.1176(20)$ & $\eta_{ct} = 0.47(4)$ \hfill \cite{HerrlichNierste} \\
$m_c(m_c) = 1.25(9)$ GeV & $\eta_{tt} = 0.5765(65)$ \hfill \cite{BurasJaminWeisz} \\
$M_t = 172.6(1.4)$ GeV\symbolfootnote[1]{The ${\rm \ov{MS}}$ mass value 
$m_t(m_t) = 162.7(1.3)$ is derived using \cite{RunDec}.} \hfill \cite{CDF-D0-top} 
                     & $F_K = 0.1561(8)$ GeV \hfill \cite{Flavianet}\\
\cline{1-1}
& \\
[-0.25cm]
$M_{B_d} = 5.2795(5)$ GeV & $M_{K^0} = 0.49765$ GeV \\
$M_{B_s} = 5.3661(6)$ GeV & $\D M_K = 0.5292(9) \cdot 10^{-2}/{\rm ps}$ \\
$\D M_d = 0.507(5)/{\rm ps}$ & $|\eps_K| = 2.232(7) \cdot 10^{-3}$ \\
$\D M_s = 17.77(12)/{\rm ps}$ \hfill \cite{CDF-DMs} & $\keps = 0.92(2)$ \\
$\xi_s = 1.21(6)$ \hfill \cite{Becirevic-CKM03,Hashimoto-ICHEP04,HPQCD-fB,Tantalo-CKM06} & $\phi_\eps = 43.51(5)^\circ$ \\
\hline
\end{tabular}
}
\caption{Input parameters. Quantities lacking a reference are taken from \cite{PDG}.}
\label{tab:input}
\end{table}

\section{T\lc{hree new-physics scenarios}}\label{sec:3NP}

Next we note that the most updated values for all the parameters on the r.h.s. of eq. (\ref{epsprop}), with 
exception of $\sin 2 \beta$, are lower with respect to previous determinations. Notably, the central value of the most 
recent estimate of $\bk$ \cite{DJAntonio} is lower by roughly 9\%, with a similar effect due to the $\keps$ factor 
(see eq. (\ref{keps})). One can then investigate whether the value of $\sin 2 \beta$ required to accommodate $|\eps_K|$ 
within the SM may be too high with respect to the $\sin 2 \beta$ determination from $B_d$ physics, as already investigated 
in \cite{LunghiSoni-Bd} for $\keps = 1$. Here we would like to emphasize that, more generally, this could entail the presence 
of a new phase either dominantly in the $B_d$ system or respectively in the $K$ system, or, alternatively, of two smaller phases in 
both systems, defining in turn three NP scenarios. Addressing the significance of either scenario crucially depends on the errors 
associated with the theoretical input entering the $\eps_K^{\rm SM}$ formula. We will come back to this point quantitatively in the 
discussion to follow, where all the present uncertainties are taken into account.

\begin{figure*}[t]
\begin{center}
\includegraphics[width=0.45 \textwidth]{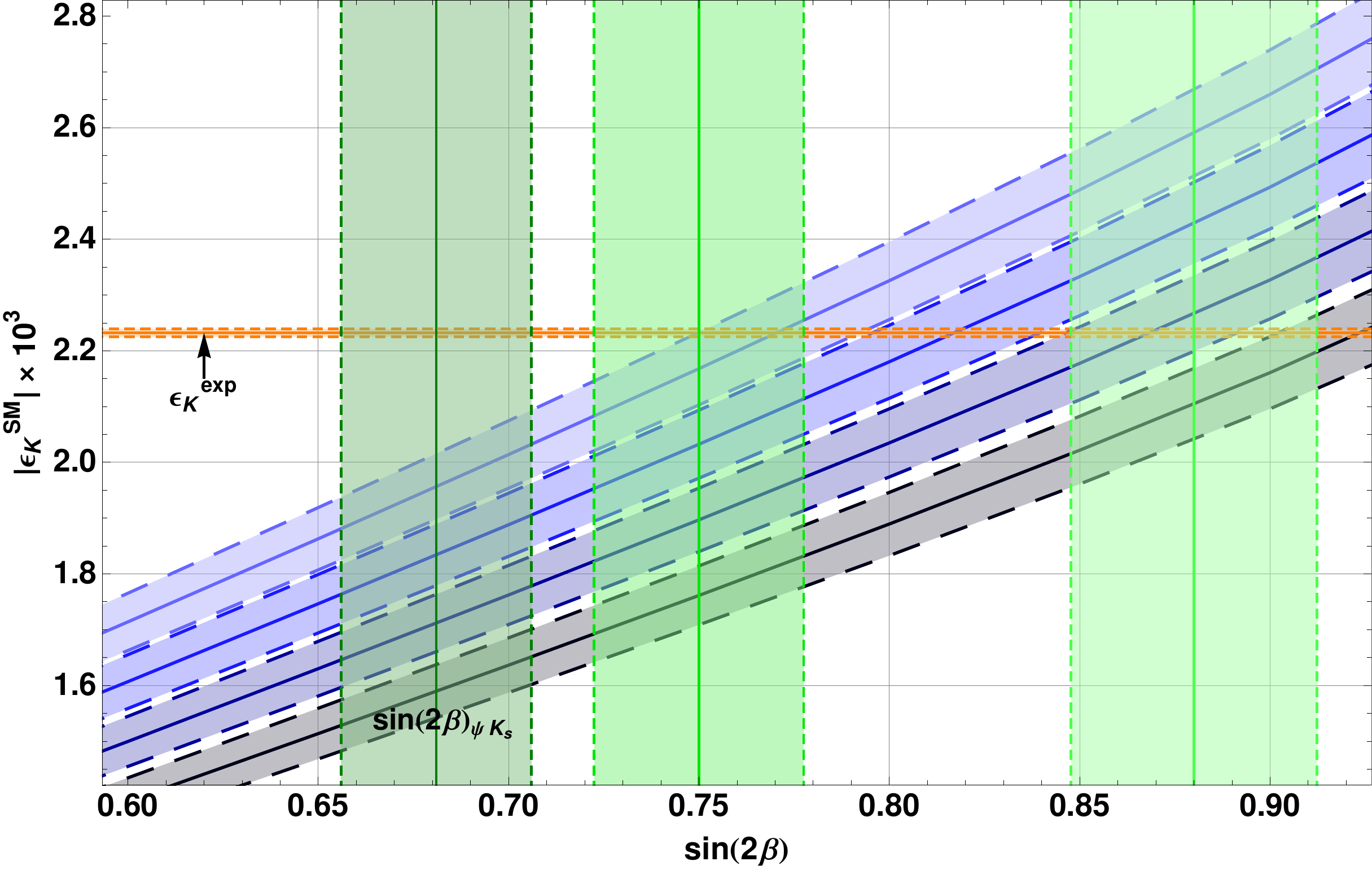}
\includegraphics[width=0.45 \textwidth]{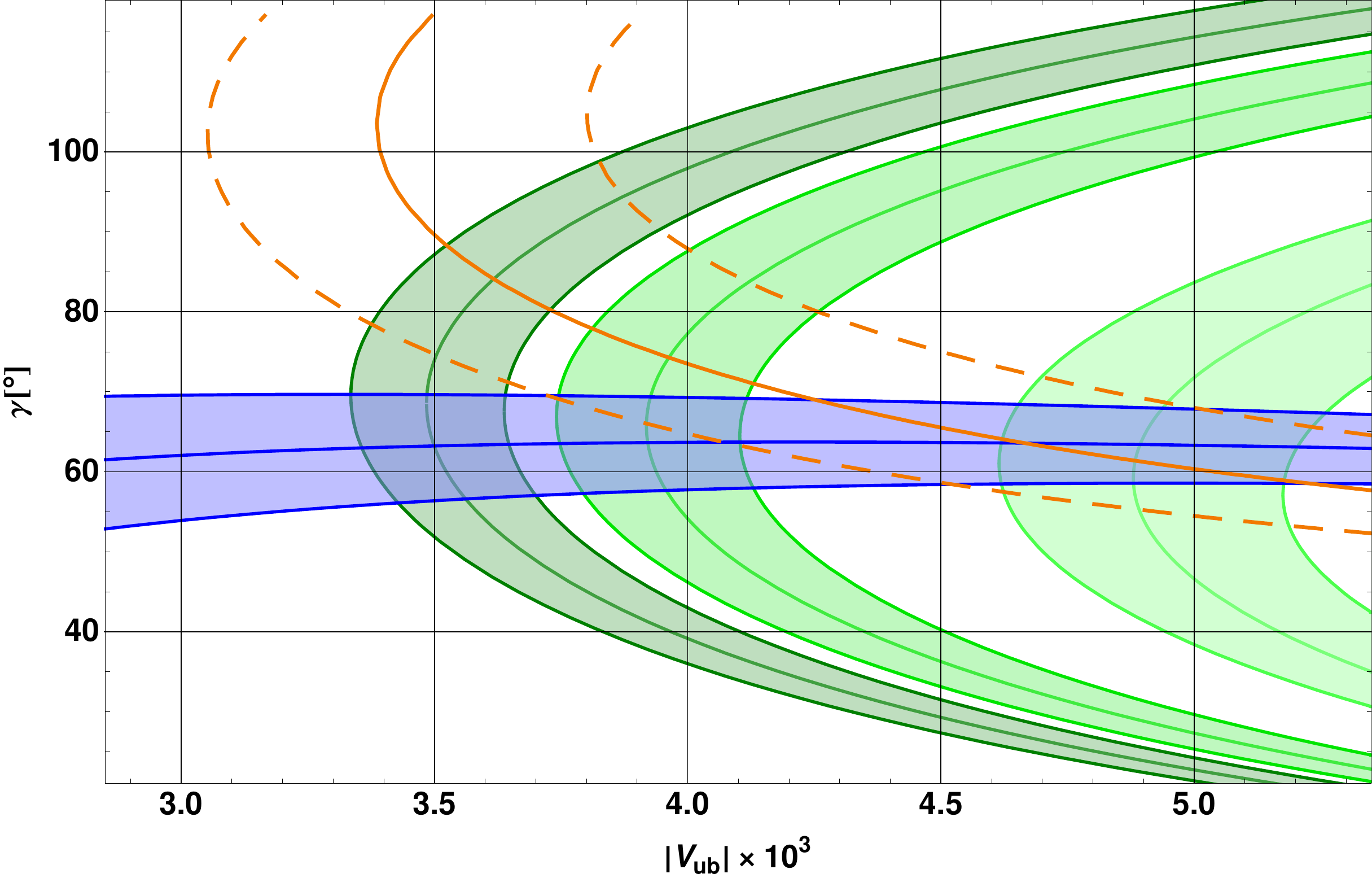}
\end{center}
\caption{\small\sl Left panel: $|\eps_K^{\rm SM}|$ vs. $\sin 2 \beta$ with only $\bk$ errors included. 
$\bk \in \{0.65, 0.70, 0.75, 0.80\} \pm 3$\% are shown as blue areas (darker to lighter). Vertical green areas display 
$\sin 2 \beta \in \{ 0.681, 0.75, 0.88 \} \pm 3.7$\% (see text). 
Right panel: $\gamma$ vs. $|V_{ub}|$ for $\sin 2 \beta \in \{0.681,0.75,0.88\} \pm 3.7$\% (green areas). 
Displayed in blue is the area corresponding to $R_t = R_t^{\rm SM}$, while orange lines represent the contours of $\eps_K^{\rm exp}$.}
\label{fig:epsK_vs_s2b_noerrors}
\end{figure*}

However, since these uncertainties in the input do not yet allow clear-cut conclusions, we would like to first illustrate the 
three just mentioned NP scenarios by setting all input parameters except $\bk$ at their central values. This would correspond to 
the hypothetical situation in which all the input, including the CKM parameters, were controlled with higher accuracy than $\bk$, 
for which we assume a 3\% uncertainty. In fig. \ref{fig:epsK_vs_s2b_noerrors} (left panel) we then show $|\eps_K^{\rm SM}|$ as a 
function of $\sin 2 \beta$ for $\bk \in \{0.65, 0.70, 0.75, 0.80\} \pm 3$\%.
The vertical ranges centered at $\sin 2 \beta \in \{ 0.681, 0.75, 0.88 \}$, with a relative error chosen at 3.7\% as in the 
$\sin 2 \beta_{\psi K_s}$ case, define the scenarios in question. 
The horizontal range, representing the experimental result for $\eps_K$, shows that $\sin 2 \beta \approx \sin 2 \beta_{\psi K_s}$ 
would require NP in $\eps_K$ in order to fit the data, unless $\bk \gtrsim 0.85$.
Conversely, in the last scenario, as considered in \cite{LunghiSoni-Bd}, no NP is required to fit the data on $\eps_K$, even for 
$\bk \approx 0.65$. In this case, however, the discrepancy with respect to the 
$\sin 2 \beta_{\psi K_s}$ determination reveals the need for a NP phase in the $B_d$ system around $-9^\circ$.
In table \ref{tab:scenarios} we report indicative values for various quantities of interest obtained from the scenarios shown in 
fig. \ref{fig:epsK_vs_s2b_noerrors} (left panel). In particular, values for $|\eps_K^{\rm SM}|$ are shown for 
$\hat B_K= \{0.7,0.8\}$. In giving the result for $S_{\psi\phi}$ we set $\phi_d=\phi_s$ (see discussion below).
\begin{table}[t]
\center{
\begin{tabular}{|r|ccc|}
\hline
 & \multicolumn{3}{c|}{$\sin 2 \beta$} \\
 & ~~0.681~~ & ~~0.75~~ & ~~0.88~~ \\ 
\hline
 & \multicolumn{3}{c|}{}\\
 [-0.2cm]
\multirow{2}{*}{$10^3 \cdot |\eps_K^{\rm SM}| \left \lbrace \begin{array}{l} \bk = 0.7 \\[0.02cm] \bk = 0.8 \end{array} \right.$} 
 & 1.71 & 1.90 & 2.27 \\
[0.1cm]
 & 1.96 & 2.17 & 2.59 \\
[0.1cm]
$\phi_d[^\circ]$ & 0 & $-2.8$ & $-9.4$ \\
[0.1cm]
$S_{\psi \phi}$ & 0.04 & 0.14 & 0.36 \\
[0.1cm]
$10^3 \cdot |V_{ub}|$ & 3.50 & 3.92 & 4.90 \\
[0.1cm]
$\ga[^\circ]$ & 63.5 & 64.0 & 63.9 \\
\hline
\end{tabular}
}
\caption{Indicative values for various quantities of interest in the scenarios represented in the left panel of 
fig. \ref{fig:epsK_vs_s2b_noerrors} (see also text).}
\label{tab:scenarios}
\end{table}
We observe that values of $\bk$ in the ballpark of 0.7 would imply a NP correction to $|\eps_K^{\rm SM}|$ exceeding $+20$\%,
which should be visible if the input parameters could be controlled with, say, 2\% accuracy.

The above discussion, and the scenarios in table \ref{tab:scenarios}, assume that the UT side $R_t$ be equal to its SM value 
(see eq. (\ref{Cd=Cs})) and imply $\gamma$ not larger than around $65^\circ$.
Figure \ref{fig:epsK_vs_s2b_noerrors} (right panel) shows the correlation existing for fixed $\sin 2 \beta$ between $\gamma$ and 
$|V_{ub}|$ (or, equivalently, the side $R_b$ \cite{ABG}). From the figure one can note that, if $\gamma$ 
from tree-level decays turns out to be larger than the values in table \ref{tab:scenarios}, consistency of $\sin 2 \beta$ with 
eq. (\ref{SpsiKs}) can be recovered by increasing the side $R_t$ with respect to the SM value (thus shifting the blue area 
in the figure upwards). As one can see from the same figure, this would also accommodate $\eps_K$, since an upward shift in 
$R_t$ from NP corresponds to $C_s > C_d$ (cf. eqs. (\ref{Rt})-(\ref{epsprop})), and could come in particular from $C_d < 1$, as 
$\D M_d$, in contrast to $\D M_s$, is directly sensitive to $R_t$.

\begin{figure*}[th]
\begin{center}
\includegraphics[width=0.45 \textwidth]{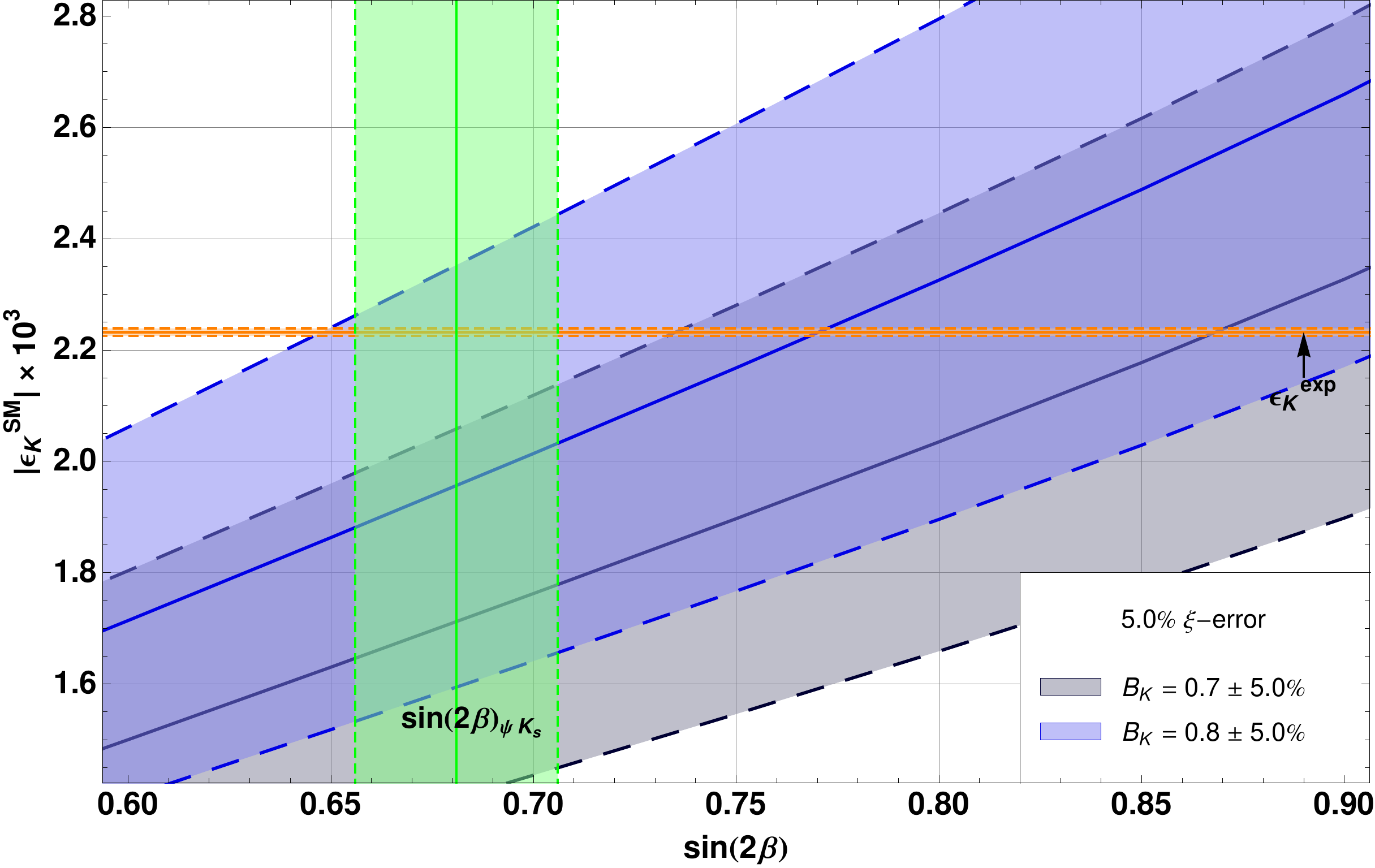}\hspace{0.5cm}
\includegraphics[width=0.45 \textwidth]{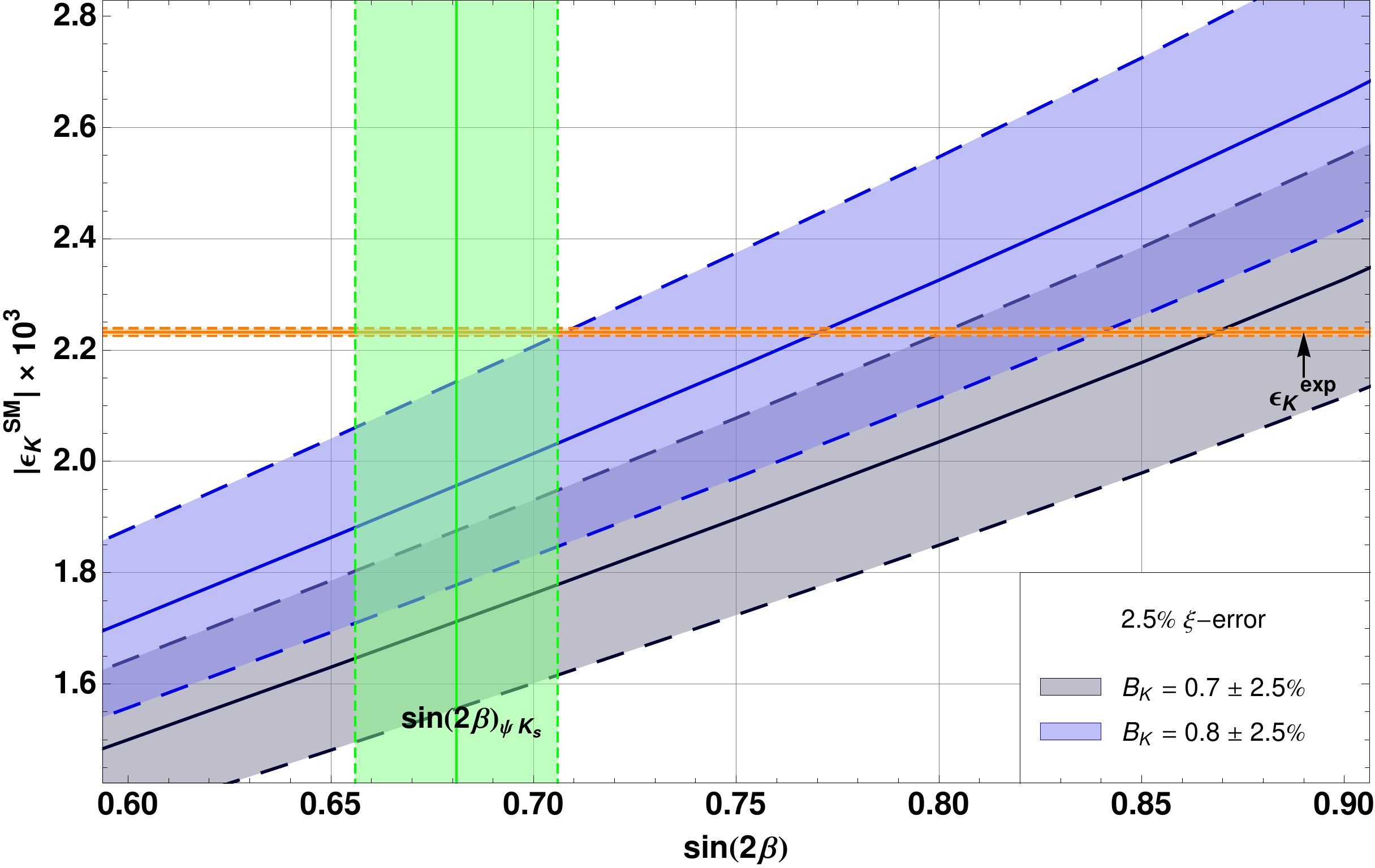}
\end{center}
\caption{\small\sl $|\eps_K^{\rm SM}|$ vs. $\sin 2 \beta$ with inclusion of all input uncertainties. Left panel assumes present 
$\bk$ and $\xi_s$ errors, whereas right panel shows the situation with errors on both quantities shrunk to 2.5\%.}
\label{fig:epsK_vs_s2b}
\end{figure*}

\begin{figure}[hb]
\begin{center}
\includegraphics[width=0.45 \textwidth]{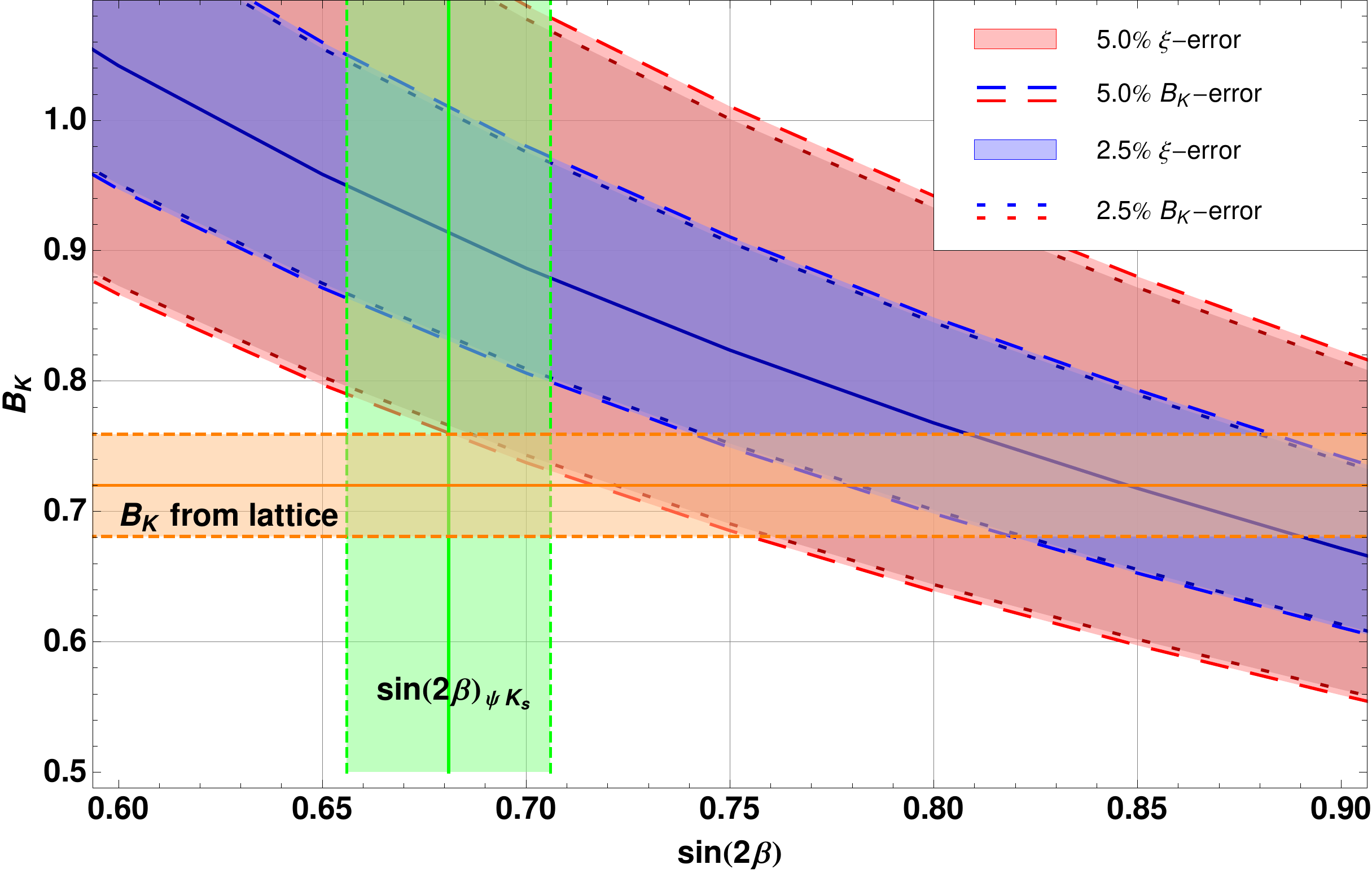}
\end{center}
\caption{\small\sl $\bk$ ranges compatible with the experimental $\eps_K$ result as a function of $\sin 2 \beta$. $\bk$ and/or $\xi_s$ 
are taken with present or 2.5\% uncertainties (see legend). Comparing red with blue areas one can note the role of a decrease in the 
$\xi_s$ error.}
\label{fig:BK_vs_s2b}
\end{figure}

%% Analysis including all errors
Plots analogous to that of fig. \ref{fig:epsK_vs_s2b_noerrors} (left panel), but with all present uncertainties on the input taken 
into account, are shown in figures \ref{fig:epsK_vs_s2b} and \ref{fig:BK_vs_s2b}. 
These plots are obtained by the following procedure. The $\D M_d / \D M_s$ constraint is used to solve for $\ov \rho, \ov \eta$ 
depending on the $\sin 2 \beta$ value. The range of solutions implied by the $\D M_d / \D M_s$ error (with $\ov \rho, \ov \eta$ 
highly correlated) can be translated into a range of values for $|\eps_K^{\rm SM}|$. The rest of the contributions to the $\eps_K$ 
error, mostly due to $m_c$, $m_t$, to the CKM entry $|V_{cb}|$ and to the assumed ranges for $\bk$ and $\keps$, can be 
treated as uncorrelated, and plugged in an error-propagation formula. As one can see, this procedure only assumes that $\D M_d / \D M_s$ 
be SM-like.

Figure \ref{fig:epsK_vs_s2b} confirms that the combined information of $\sin 2 \beta_{\psi K_s}$ and $|\eps_K^{\rm exp}|$ tends to prefer 
`high' values of $\bk \gtrsim 0.85$ (cf. estimate in \cite{AOV-BK}). However, use of present errors on $\bk$ and $\xi_s$ (both $\approx$ 
5\%), as in the left panel of fig. \ref{fig:epsK_vs_s2b}, impairs any clear-cut conclusion. The situation in the case of $\bk$ and $\xi_s$ 
errors hypothetically halved can be appreciated from the right panel of the same figure, where actually a large part of the improvement 
is driven by the shrinking in the $\xi_s$ error, allowing a better determination of $\ov \rho, \ov \eta$. Therefore an alternative or 
complementary strategy to an improvement in $\xi_s$ would be a major advance in the angle $\gamma$ through tree-level decays.

Finally, as an alternative viewpoint on the above facts (in particular on the role of the $\bk$ and $\xi_s$ errors), 
figure \ref{fig:BK_vs_s2b} displays, as a function of $\sin 2 \beta$, the $\bk$ range compatible with the experimental $\eps_K$ result. 
For $\sin 2\beta = \sin 2\beta_{\psi K_s}$ the required $\bk$ agrees well with the one found in \cite{AOV-BK}.

\mysection{$S_{\psi \phi}$ \lc{and $\sin 2 \beta$}}\label{sec:B}

\noi As a last case, we would like to focus on the possibility that NP contributions to $\eps_K$ be negligible, as assumed 
in \cite{LunghiSoni-Bd} and in scenario 3 discussed in the previous section. 
As one can infer from the above considerations, this would favor values of $\sin 2 \beta \gtrsim 0.80$, implying the presence of 
a sizable new phase in $B_d$ mixing with a possible correlation with the $B_s$ system, which we discuss next.

Let us start with the $B_s$ mixing phase $\beta_s^{\rm full}$, eq. (\ref{betadsdefs}), using the information from \cite{UTfit-Bs}.
In the notation of our eqs. (\ref{parameterization})-(\ref{betadsdefs}), the range for the NP phase $\phi_s$ at 95\% probability 
is found to be
\beqn
&&\phi_s \in [-30.45,-9.29]^\circ \cup [-78.45,-58.2]^\circ~,\nn \\
&&\mbox{corresponding to } S_{\psi \phi} \in [0.35, 0.89].
\label{phis}
\eeqn
Assuming generic NP, the SM contribution to the phase amounts instead to \cite{UTfit-Bs} $\beta_s = -1.17(11)^\circ$, where to estimate 
the error we have simply propagated that on $\sin 2 \beta_s$.

Let us now compare these findings with the $B_d$ case. If a NP phase contributes to the mixing amplitude, the CP asymmetry in 
$B_d \to \psi K_s$ measures the quantity $\beta_d^{\rm full}$ (see eq. (\ref{betadsdefs})). Then, one can extract information on the 
NP phase $\phi_d$, provided the SM phase $\beta$ is estimated in some other way. An example is the determination of 
ref. \cite{LunghiSoni-Bd}, where the main assumptions are the absence of NP in the ratio $\D M_d / \D M_s$ and in $\eps_K$ 
(as we are supposing in the present scenario). Using the CKMfitter package \cite{CKMfitter}, we find
\beqn
\sin (2 \beta) = 0.88^{+0.11}_{-0.12}~,
\label{betaLS}
\eeqn
where we have used the $\bk$ result from ref. \cite{DJAntonio} and the $\keps$ factor in table \ref{tab:input} and, similarly to 
ref. \cite{LunghiSoni-Bd}, we have treated all the input errors as Gaussian. The result in eq. (\ref{betaLS}) is compatible with that 
of \cite{LunghiSoni-Bd}: in particular, the inclusion of the $\keps$ correction pushes the $\sin 2 \beta$ determination further upwards, 
even if its associated error introduces an additional uncertainty in the $\eps_K$ evaluation.

If the high value implied by eq. (\ref{betaLS}) for $\beta$ were indeed correct, this would indicate the presence of a {\em negative} 
NP phase in the $B_d$ system, with absolute value of O(10$^\circ$). Quite interestingly, the solutions found in \cite{UTfit-Bs}
for the NP phase in the $B_s$ system (see eq. (\ref{phis})) go in the same (negative) direction and the lowest solution is also compatible 
with $\approx - 10^\circ$.

\begin{figure}[t]
\begin{center}
\includegraphics[width=0.45 \textwidth]{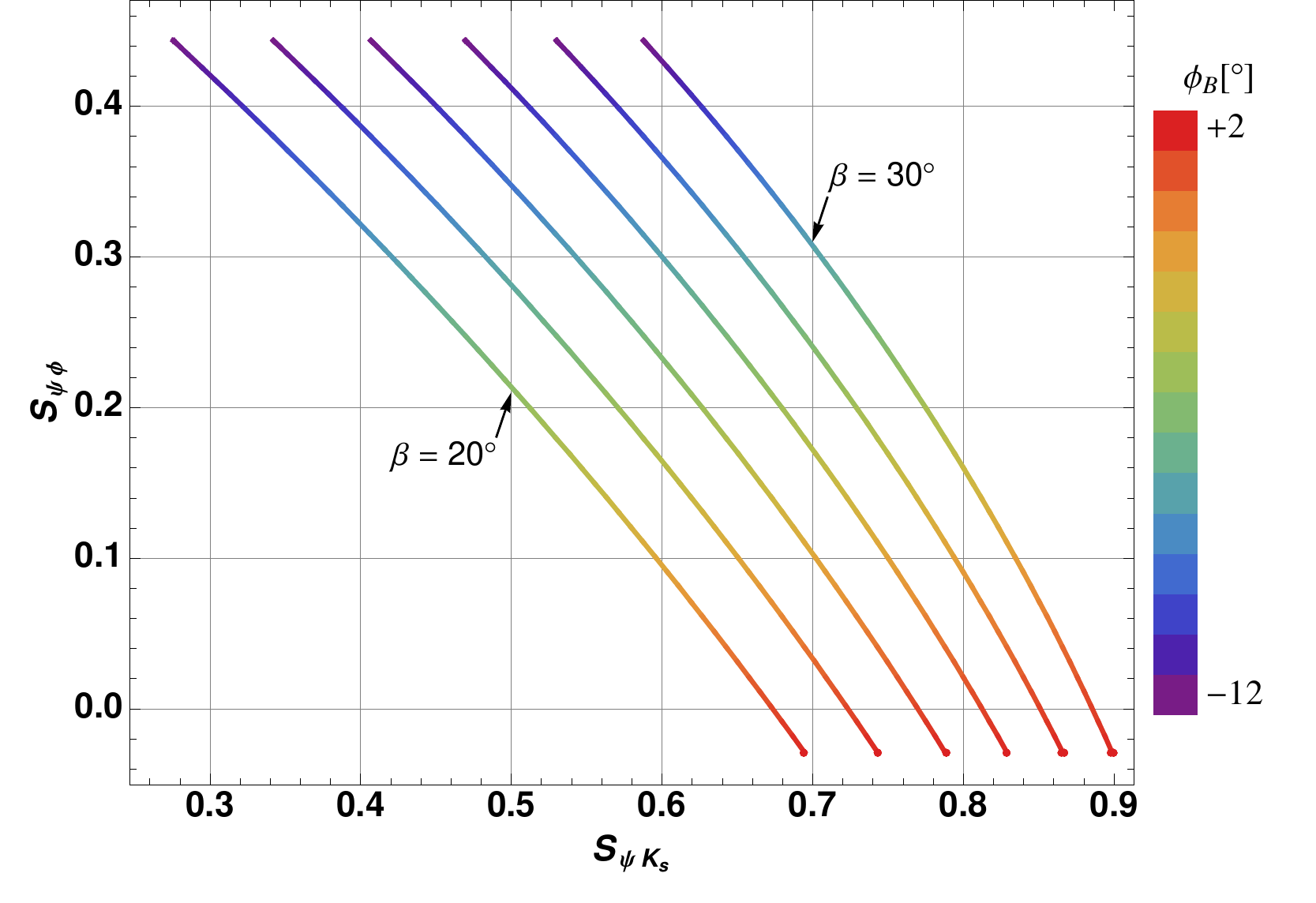}
\end{center}
\caption{\small\sl CP asymmetry $S_{\psi \phi}$ as a function of $S_{\psi K_s}$ for a common NP phase 
$\phi_B \in [-12, +2]^\circ$.}
\label{fig:S_vs_S}
\end{figure}

\begin{figure}[t]
\begin{center}
\includegraphics[width=0.45 \textwidth]{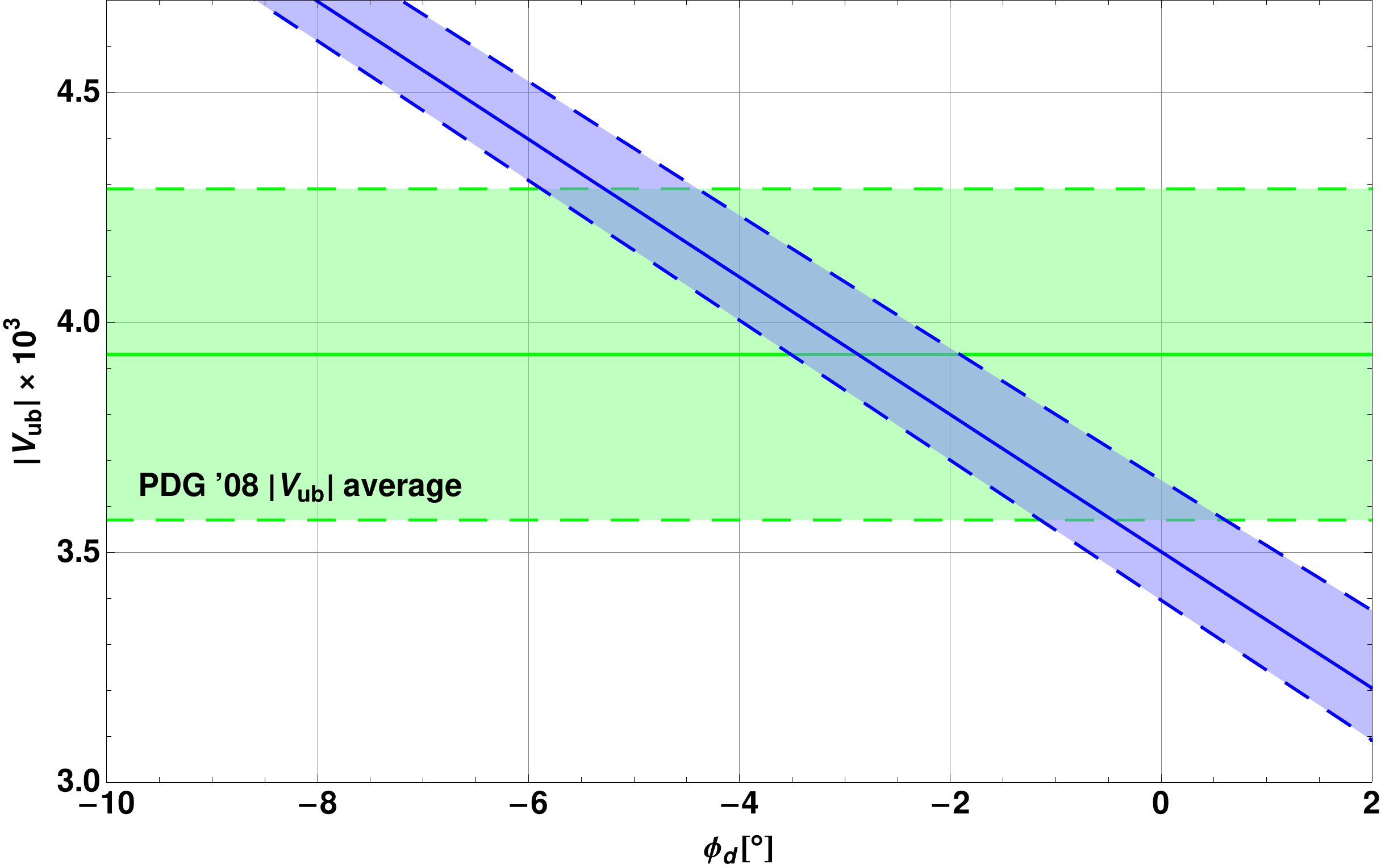}
\end{center}
\caption{\small\sl $|V_{ub}|$ ranges implied by a given NP phase in the $B_d$ system, $\phi_d$. The green band displays the most recent $|V_{ub}|$ 
average quoted in the PDG \cite{PDG08}.}
\label{fig:Vub_vs_phid}
\end{figure}

One is then tempted to envisage a scenario characterized by a significant NP phase roughly equal in both $B_d$ and $B_s$ systems, i.e.
\beqn
\phi_B = \phi_d \approx \phi_s \approx - 9^\circ 
\Rightarrow
\left \lbrace 
\begin{array}{l}
\beta_{\psi K_s} < \beta \approx 30^\circ \\
S_{\psi \phi} \approx 0.4
\end{array}
\right.
\label{scenario3}
\eeqn
with no NP in the $K$ system. The interesting aspect of this scenario is the correlation between new CP violation in the $B_d$ 
and $B_s$ systems. In the limiting case of exact equality between the NP phases in the two sectors, we show in figure \ref{fig:S_vs_S}
the predicted $S_{\psi \phi}$ as a function of $S_{\psi K_s}$ (see eqs. (\ref{SpsiKs})-(\ref{Spsiphi}) for the definitions).
If improvements on the $\sin 2 \beta$ determination should indicate a large figure like eq. (\ref{betaLS}) and $S_{\psi \phi}$ were measured 
as large as $0.4$, this could be a hint in favor of this scenario. On the other hand, the scenario in eq. (\ref{scenario3}) seems to be problematic 
with regards to the implied $|V_{ub}|$ value. As seen already in the right panel of fig. \ref{fig:epsK_vs_s2b_noerrors}, the value of $|V_{ub}|$ 
is generically larger than the present exclusive result. To address this issue, we plot in figure \ref{fig:Vub_vs_phid} the $|V_{ub}|$ range 
implied by a given NP phase $\phi_d$. We note that, since $|V_{ub}|$ is determined from the side $R_b$, its error depends mostly on the $|V_{cb}|$ 
uncertainty, and is estimated through the propagation formula. On the other hand, for fixed $\sin 2 \beta$, $|V_{ub}|$ depends only very weakly 
on the error due to the $\ov \rho, \ov \eta$ determination, as expected.

From figure \ref{fig:Vub_vs_phid} and table \ref{tab:scenarios} it is evident that $\phi_d \approx - 9^\circ$ would imply 
$|V_{ub}| \approx 4.9 \times 10^{-3}$, which is even higher than the inclusive averages in \cite{Barberio}. For comparison, the most recent
combination of the inclusive and exclusive $|V_{ub}|$ determinations quoted in the PDG \cite{PDG08}, namely
\beqn
|V_{ub}| = (3.93 \pm 0.36) \times 10^{-3}~,
\label{VubINEX}
\eeqn
is reported in figure \ref{fig:Vub_vs_phid} as a green band, and can be seen to be compatible with no phase. Results similar to eq. (\ref{VubINEX})
can be found in \cite{Neubert-LP07}.

Therefore, assuming that $|V_{ub}| \lesssim 4 \times 10^{-3}$ and that $S_{\psi \phi}$ should be confirmed as large as implied by 
eq. (\ref{phis}), the middle scenario presented in the previous section, characterized by smaller NP effects in both the $B_d$ and $K$ 
sectors, would be a more plausible possibility. In this case, the NP phases in the $B_d$ and $B_s$ systems would be (mostly) uncorrelated 
with each other. In fact, in the case of exact correlation (see figure \ref{fig:S_vs_S}), $\beta \approx 26^\circ$, corresponding to scenario 2, 
would imply $S_{\psi \phi} \lesssim 0.2$. As for the $K$ system, ascertaining the presence of NP would require a leap forward in the input errors, 
$\bk$ and $|V_{cb}|$ in the first place.

\mysection{C\lc{onclusions}}\label{sec:concl}

\noi In the present paper we have pointed out a possible inconsistency between the size of CP violation in $K - \ov K$ and $B_d - \ov B_d$ 
mixing within the SM. The recent decrease in $\bk$ from lattice \cite{DJAntonio,JLQCD-BK} and the inclusion of the suppression factor $\keps$ 
in the formula for $\eps_K$ are mostly responsible for this finding. Such an inconsistency has been already noted 
in \cite{LunghiSoni-Bd}, but we differ from that paper as we do not assume the absence of NP in $\eps_K$. Moreover, in \cite{LunghiSoni-Bd}
$\keps \simeq 0.92$ has not been taken into account.

Under the single assumption that $\D M_d/\D M_s$ be unaffected by NP, the general pattern of correlations between CP violation in 
the $K - \ov K$ and $B_d - \ov B_d$ systems is as follows:

\begin{itemize}

\item In the absence of new CP violation  in the $B_d$ system, the measured size~of $S_{\psi K_s}$ implies $\eps_K$ with a central value 
as much as 20\% below the data, hinting at NP in $K - \ov K$ mixing.

\item In the absence of new CP violation in $K - \ov K$ mixing, the size of the measured value of $\eps_K$ implies $\sin 2\beta$ 
by 10-20\% larger \cite{LunghiSoni-Bd} than $S_{\psi K_s}$, so that a negative new phase $\phi_d$ is required in order~to fit the experimental 
value of $S_{\psi K_s}$.

\item Since $\phi_d$ can reach O($-10^\circ$), the limiting case of a new phase roughly equal in both $B_d$ and $B_s$ systems allows an
enhancement of the asymmetry $S_{\psi \phi}$ by roughly an order of magnitude with respect to its SM value. This could then explain, at least
to a first approximation, the effect found in \cite{UTfit-Bs}.
\end{itemize}

If, on the other hand, one allows for contributions of NP to $\D M_d/\D M_s$, so~that $R_t$ is increased with respect to its SM value, one 
can~remove the discrepancy between the two systems, provided $R_t$ is increased by, say, 10-15\%. This would require, for instance, a 
destructive interference between SM and NP contributions to $\D M_d$ -- i.e., recalling eq. (\ref{Rt}), $C_d < 1$ -- and would automatically 
increase also $\gamma$.

Finally, we would like to emphasize that our results are very sensitive to the used value of $V_{cb}$, as can be anticipated from 
eq. (\ref{epsprop}). Therefore, in addition to an accurate calculation of $\bk$ and $\xi_s$, a very precise determination of $V_{cb}$ is 
required in order to fully exploit the power of the $\eps_K$ constraint on NP.

We hope that the results and the plots in our paper will help to monitor the developments in the field of $\D F=2$ transitions in the coming 
years, when various input parameters and the data on CP violation in $b \to s$ transitions will steadily improve.

\begin{acknowledgments}
\noi We thank Uli Nierste for discussions related to section \ref{sec:K}, Monika Blanke for critical comments on the manuscript and Federico Mescia 
for kind feedback on input parameters from Flavianet. We also thank Alexander Lenz and Paride Paradisi for useful discussions.
This work has been supported in part by the Cluster of Excellence ``Origin and Structure of the Universe'' and by the German Bundesministerium 
f{\"u}r Bildung und Forschung under contract 05HT6WOA. D.G. also warmly acknowledges the support of the A. von Humboldt Stiftung.
\end{acknowledgments}

\mysectionA{A\lowercase{ppendix}: E\lowercase{stimate of the parameter $\ovkeps$}}
\setcounter{equation}{0}
\renewcommand{\theequation}{A.\arabic{equation}}

\noi A rough estimate of the factor $\ovkeps$, discussed at the beginning of section \ref{sec:K}, can be obtained as follows. Starting from 
the general formula for $\eps_K$ in eq. (\ref{epsexact}), one finds
\beqn
\ovkeps \simeq 1 + \frac{\xi}{\sqrt 2 |\eps_K|} \equiv 1 + \D_\eps~,
\eeqn
where terms of O($\xi^2$) on the r.h.s. have been neglected. Then $\D_\eps$
can in principle be extracted from the analyses of $\eps^\prime / \eps$. One has \cite{BurasLesHouches}
\beqn
\frac{\eps^\prime}{\eps} = - \omega \D_\eps(1 - \Omega)~,
\eeqn
where $\omega = {\rm Re} A_2 / {\rm Re} A_0 = 0.045$ and $\Omega$ summarizes the isospin-breaking corrections, that are dominated by 
electroweak penguin contributions. It is well known that $\Omega > 0$ in the SM and in most known SM extensions. Therefore, setting 
$\Omega = 0$ and using the experimental value for $\eps^\prime / \eps = 1.66(26) \times 10^{-3}$ \cite{PDG}, one finds 
\beqn
\D_\eps = - \frac{1}{\omega} \frac{\eps^\prime}{\eps} = (-3.7 \pm 0.6) \times 10^{-2}~,
\label{Deps1}
\eeqn
which is compatible with \cite{AOV-epsK,AOV-BK}. This value can be considered as a plausible lower bound on $|\D_\eps|$.

However, it is well known that $\Omega$ cannot be neglected, but the evaluation of this quantity is subject to significant hadronic 
uncertainties, although, as discussed in ref. \cite{BurasJamin}, these uncertainties appear to~be smaller than in $\xi$ itself. 
We recall that $\xi$ and $\Omega$ are dominated by QCD penguin and electroweak penguin operators respectively, and the evaluation of 
$\xi$ and $\Omega$ requires the knowledge of their hadronic matrix elements.

One method \cite{FNAL-report-2002} is to evaluate $\Omega$ and extract $\D_\eps$ from $\eps^\prime / \eps$. From the analysis 
of \cite{BurasJamin}, that combined various non-perturbative approaches, we find $\Omega = 0.4 \pm 0.1$ in the SM. 
Yet, one has to remember that $\Omega$ is sensitive to NP contributions, in contrast with $\D_\eps$, whose NP sensitivity turns out to be 
much smaller. For this reason we have also calculated $\D_\eps$ directly in the large $N_c$ approach \cite{BardeenBurasGerard-2}. Both 
routes give
\beqn
\D_\eps \simeq -6 \times 10^{-2}~.
\label{Deps2}
\eeqn
Calculations (\ref{Deps1}) and (\ref{Deps2}) and the fact that the SM estimate of $\eps^\prime / \eps$ in the large $N_c$ approach agrees
well with the data \cite{LHT-eps} drive us to the estimate
\beqn
\ovkeps \approx 0.94 \pm 0.02~.
\label{kepstimate}
\eeqn
This agrees well with the 6\% effect estimated in \cite{FNAL-report-2002}. The error quoted in (\ref{kepstimate}) is no more than 
a guesstimate, but we believe it to be realistic. Clearly a better calculation of $\ovkeps$ should be attempted, using e.g. lattice methods.
The result obtained in \cite{AOV-BK} through a direct calculation of $\xi$ corresponds to $\ovkeps \simeq 0.90(3)$ and implies
$\eps^\prime / \eps \approx 4.5 \times 10^{-3}$ from QCD penguins alone, roughly by a factor 3 larger than the data. Such result requires 
a very large negative electroweak penguin component for the predicted $\eps^\prime / \eps$ to agree with experiment and a certain fine 
tuning between the two contributions. Consequently we believe that eq. (\ref{kepstimate}) represents a very plausible estimate of $\ovkeps$.

\bibliography{BG}

\end{document}